\documentclass[runningheads]{svmult}

\usepackage{makeidx}   
\usepackage{graphicx}  
\usepackage{subeqnar}  
\usepackage{multicol}  
\usepackage{physprbb}  
\makeindex             

\def\lesssim{\mathrel{\hbox{\rlap{\hbox{\lower4pt\hbox{$\sim$}}}\hbox{$<$}}}}
\def\gtrsim{\mathrel{\hbox{\rlap{\hbox{\lower4pt\hbox{$\sim$}}}\hbox{$>$}}}}

\begin{document}
\title*{A Keck/Deimos Survey of Red Giant Branch Stars in the Outskirts of M31}
\toctitle{A Keck/Deimos Survey of Red Giant Branch Stars in the Outskirts of M31}
%
%
\titlerunning{A Keck/Deimos Survey of Red Giant Branch Stars in the Outskirts of M31}
%
\author{Annette M. N. Ferguson\inst{1}
\and Scott Chapman\inst{2}
\and Rodrigo Ibata\inst{3}
\and Mike Irwin\inst{4}
\and Geraint Lewis\inst{5}
\and Alan McConnachie\inst{4}}

\authorrunning{Annette M. N. Ferguson et al.}
%
%
\institute{Max-Planck-Institut f\"{u}r Astrophysik, Postfach 1317, 87541 Garching, Germany
\and California Institute of Technology, Pasadena, CA 91125, USA
\and Observatoire de Strasbourg, 11, rue de l'Universit\'{e}, Strasbourg F-67000, France
\and Institute of Astronomy, Madingley Road, Cambridge CB3 0HA, UK
\and School of Physics, University of Sydney, NSW 2006, Australia}

\maketitle              

\begin{abstract}
  We are using the DEIMOS multi-object spectrograph on the Keck~II
  10~m telescope to conduct a spectroscopic survey of red giant branch
  stars in the outskirts of M31.  To date, velocities have been
  obtained for most of the major substructures in the halo as well as
  at several positions in the far outer disk and inner halo. First
  results concerning the giant stellar stream and major axis
  substructures are presented here.
\end{abstract}

\section{Introduction}

Within the popular $\Lambda$CDM model for structure formation, massive
galaxies are built up through the merger and accretion of smaller
subsystems and through the smooth accretion of gas. Under
the assumption that at least some of the accreted satellites contain
significant stellar components, one expects observable signatures of
this process in the form of tidal streams and other stellar
inhomogeneities.  While such features have been previously found in
our Milky Way, little has been known, until recently, about their
frequency in other galaxies.  One of the main goals of the Isaac
Newton Telescope Wide-Field Camera (INT WFC) survey of M31
\cite{ibata01,ferg02,irwin04} has been to search for signatures of
satellite disruption in the outer halo of our nearest giant neighbour.

During the past four years, we have used the INT WFC to map $\approx$
40 sq. deg. (163 contiguous pointings) around M31 with 800--1200 sec
exposures in the V and {\it i} passbands
\cite{ibata01,ferg02,irwin04}.  This depth is sufficient to reach
V$\sim 24.5$, {\sl i}$\sim 23.5$, and hence probe the top three
magnitudes of the red giant branch (RGB). Our current coverage extends
to 4 degrees ($\approx$55~kpc) and 2.5 degrees ($\approx$30~kpc) along
the major and minor axes respectively.  Figure \ref{fig1} shows the
distribution of ``blue'' RGB stars across our survey area.  As already
apparent from our interim results \cite{ibata01,ferg02}, the
distribution of giant branch stars at large radius is very far from
uniform.  Many of the clumpy features visible in the stellar
distribution have effective surface brightnesses $\Sigma_{V}\gtrsim
28$ mag sq. arcsec -- a value typically unobservable by traditional
techniques but possible here due to the fact we are resolving
individual stars at the bright end of the luminosity function.

The mere existence of stellar overdensities in the outskirts of M31
indicates an active accretion history, however many important
questions remain. Are the overdensities the result of many small
accretions, or one large one?  How much of the structure is simply the
result of a warped/perturbed outer disk?  What is the relationship
between various stellar overdensities, such as a the giant tidal
stream, and M31's innermost satellites?  To address these issues, we
are pursuing a comprehensive follow-up program consisting of a large
radial velocity survey with Keck~II/DEIMOS and a detailed stellar
populations study with HST/ACS.

\section{Surveying the M31 Outer Halo with Keck/DEIMOS}
Stellar kinematics at the distance of M31 can be probed directly
through radial velocities of individual giant stars, or indirectly
through tracer populations, such as planetary nebulae (PNe).  While
PNe have been used successfully to study the kinematics of the disk
and inner halo (e.g. \cite{merr03}), the efficiency of this technique
declines rapidly in the low surface brightness outer regions. If one
assumes the canonical $\alpha_{2.5}\sim50\times10^{-9}$ PNe per unit V-band
luminosity \cite{ciard89} then $\approx50$ PNe would be
expected per square degree at $\Sigma_{\rm V} \sim 24$~mag sq. arcsec but 
only $\approx1$ would be expected within the same area at $\Sigma_{\rm V}
\sim 28$~mag sq. arcsec.

\begin{figure}[h]
\begin{center}
\includegraphics[width=8cm]{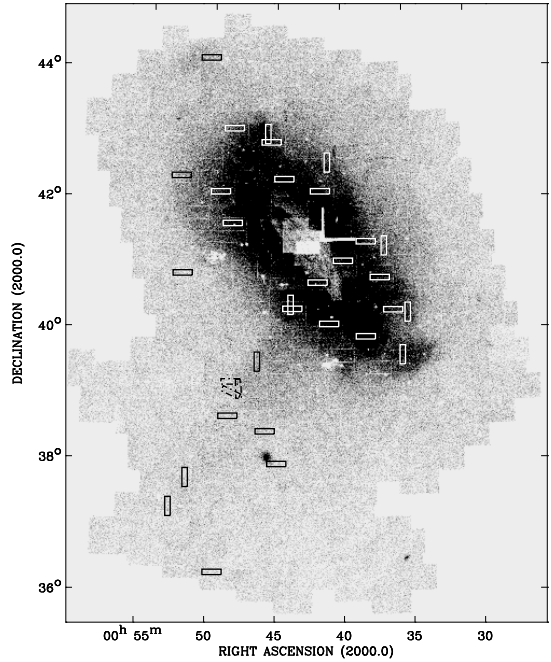}
\end{center}
\caption[]{The distribution of $``$blue'' red giant branch stars in a
  $\approx 40$ square degree (125 x 95~kpc) area centreed on M31, as
  mapped with the INT WFC.  Overlaid are the 30 Keck/DEIMOS pointings
  observed as of September 2003; these mainly target major regions of
  stellar substructure and a few locations in the far outer disk/inner
  halo.  The three DEIMOS fields of Guhathakurta et al \cite{raja04}
  are shown as dashed rectangles.}
\label{fig1}
\end{figure}

We have been conducting our radial velocity survey using the DEIMOS
multi-object spectrograph on the KeckII~10m telescope (see Figure \ref{fig1}
for the pointings observed to date).  Our strategy involves both
standard multi-slit masks, resulting in $\approx100$ targets per
$16.9'\times5.0'$ DEIMOS pointing, as well as the use of a
custom-built narrow-band filter to limit wavelength coverage and
increase multiplexing in high density regions, typically allowing 
$\sim400$ targets per mask.  We exploit the near-IR Calcium II triplet
lines ($\sim 8500$\AA) to provide information about both radial
velocities (typical uncertainty here 5--10~km/s) and metallicities.
Full details of our survey strategy and results so far are given in
\cite{ibata04,mccon04,chap04}.
 
\section{Results To Date}

\subsection{The Giant Stellar Stream}

\begin{figure}
\begin{center}
\includegraphics[angle=90,width=11cm]{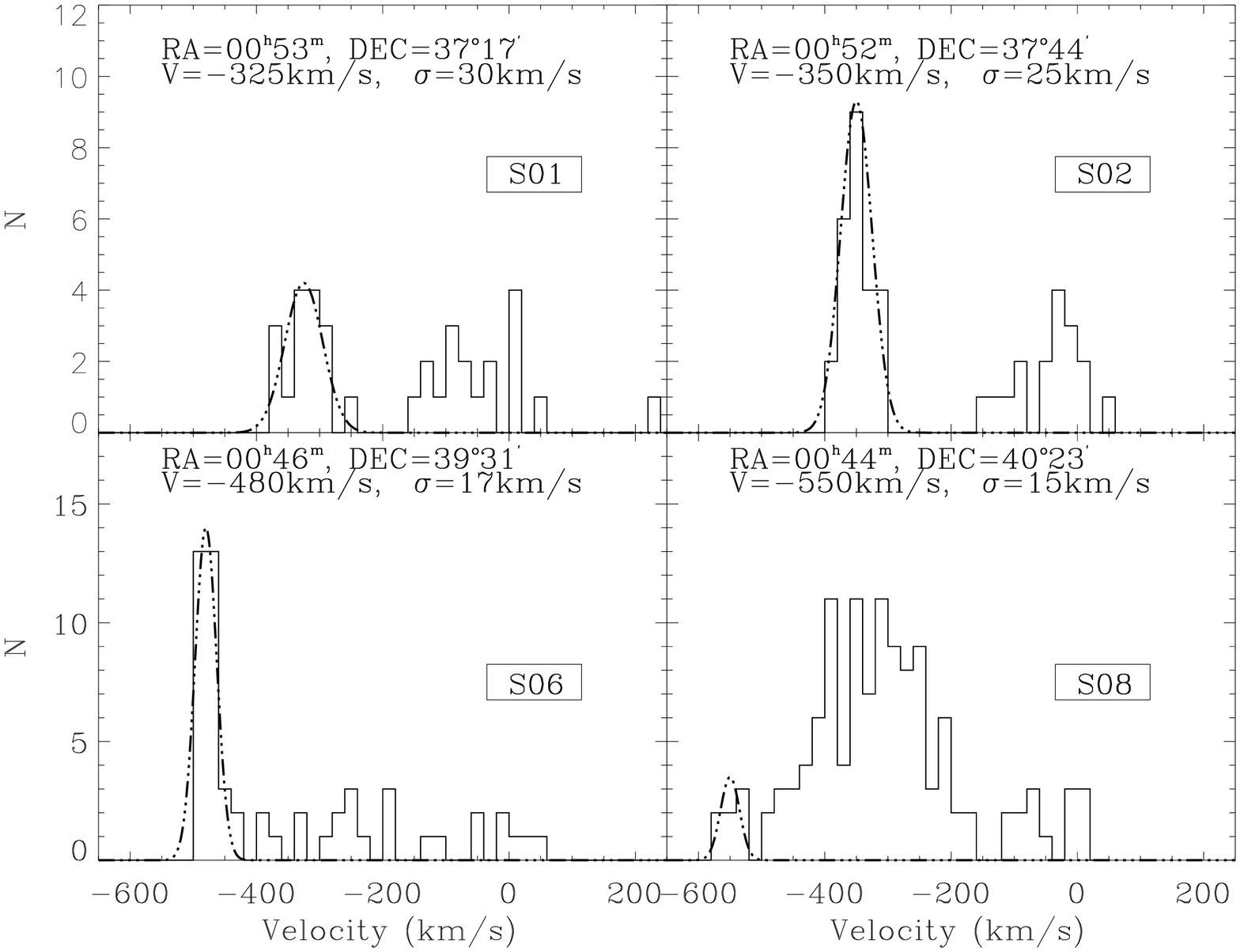}
\end{center}
\caption[]{Stellar radial (heliocentric) velocity distributions at four locations lying along the giant stellar stream (see \cite{ibata04}).  The dashed-dotted lines represent
  gaussian fits to the components attributed to the stream stars.  The
  measured velocity dispersion of the stream declines with distance to
  M31.  The systemic velocity of M31 is $-300$~km/s.}
\label{fig2}
\end{figure}

Radial velocities for stars lying at four locations along the giant
stellar stream were presented by \cite{ibata04} and used in
combination with distance data \cite{mccon03} to place constraints on the
orbit of the stream progenitor and on the total mass of M31. Figure
\ref{fig2} shows velocity histograms at these locations for stars
with good S/N and good quality cross-correlation
measurements.  Gaussians distributions have been used to describe the 
components attributed to the stream (the dominant component in all fields
significantly displaced from the inner halo (i.e. S01, S02, S06)) and the
centroids and widths are indicated.  Note that considerably more stars
are now present in S08 than in \cite{ibata04} due to a reanalysis of
the original data.  A smooth gradient in both radial
(heliocentric) velocity and line-of-sight velocity dispersion
(uncorrected for the instrumental errror) is apparent along the stream.
The measurements of \cite{raja04} for fields lying between S02
and S06 are fully consistent with these trends.  The southern
extent of the stream is essentially at rest with respect to M31
($\Delta{V_{\rm M31}}\sim-25$~km/s) and has the largest observed velocity 
dispersion ($\sigma_{v}\sim30$~km/s) -- these observations suggest
this location is close to apocentre \cite{ibata04,font04}.

The best-fitting progenitor orbit (in the best-fitting potential, see
\cite{ibata04}) is highly radial and viewed close to edge-on.  It
passes near the centre of M31 before looping around to a position
north-east of the galaxy centre (also supported by detailed studies of
the stellar populations in these parts \cite{ferg02,ferg04}).  Before
velocity data were available, a prime candidate for the progenitor was
the compact dE, M32, which projects directly onto the stream, has a
comparable line-of-sight distance and shares a similarily high
metallicity. This association no longer appears likely however, since
the radial velocity of M32 ($\Delta{V_{\rm M31}}=+100$~km/s) is
inconsistent with the expected velocity of the stream at this position
in the current orbital phase ($\Delta{V_{\rm M31}}\sim-280$~km/s).
Similar arguments make an association with NGC~205, M31's second
closest luminous satellite, equally unlikely.

\subsection{Substructures Along the Major Axis}

\begin{figure}
\begin{center}
\includegraphics[angle=90,width=10cm]{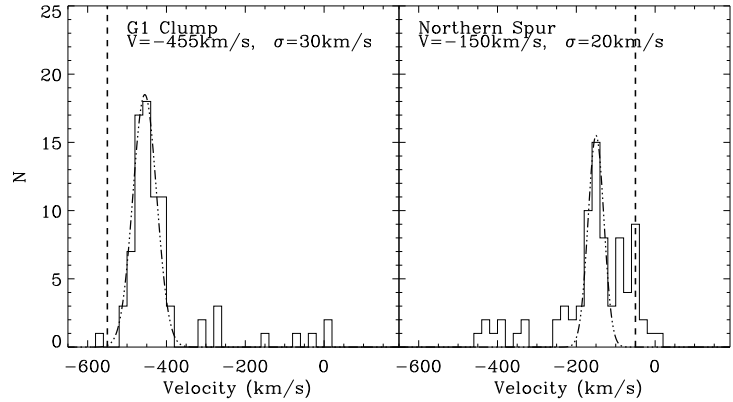}
\end{center}
\caption[]{Radial velocity distributions for regions lying close to the north-eastern and
  south-western major axes. The dashed-dotted lines represent gaussian
  fits to the dominant component in these fields, while the vertical
  dashed lines indicate simple expectations for the disk rotation
  velocity at these locations along the major axis. }
\label{fig3}
\end{figure}

Significant substructure has been identified along both the
north-eastern and south-western major axes (Figure \ref{fig1}),
termed the ``northern spur'' and the ``G1 clump'' respectively.  These
stellar overdensities are unlikely to be related to each other -- for
e.g. representing debris from a satellite orbiting within the disk
plane -- given the different colours exhibited by their constituent
stellar populations \cite{ferg02,ferg04}.  Figure \ref{fig3} presents
radial velocity histograms for these regions constructed from stars with 
good S/N measurements.  The G1 clump, located at a radius of $\sim
35$~kpc, exhibits a clear velocity peak at $-455$~km/s (see also
\cite{reitzel04}). On the other hand, the G1 globular cluster, which
is projected near the edge of the overdensity, has a sufficiently
different radial velocity ($-331$~km/s),  making it unlikely that the
two are related. The northern spur region, located at a radius of $\sim
25$~kpc, exhibits a clear velocity peak at $-150$~km/s.  While it is
tempting to associate the major axis substructures with the outer
disk, we note that simple expectations for the disk velocity (assuming
V$_{rot}=250$~km/s, based on HI observations \cite{braun91}) are $-550$~km/s and
$-50$~km/s respectively for the clump and the spur -- velocities which
differ from those observed by $\sim 100$~km/s (Figure \ref{fig3}). 

\section{Summary}

Our spectroscopic survey of RGB stars in the outkirts of M31 aims to
constrain the nature and origin of the stellar substructure observed
in these parts, and to quantify the kinematic and metallicity
structure of the far outer disk and inner halo. Our results so far
have enabled M32 and NGC~205 to be ruled out as the progenitors of
the giant stellar stream; instead, they suggest an orbit which
connects the stream to the diffuse overdensity located north-east of
the galaxy centre. The kinematics of the substructure observed along
the major axes defy easy interpretation at present.  Future
observations of radial velocities in unperturbed regions of the M31
outer disk will be crucial.

\end{document}